\documentstyle[amssymb,preprint,aps]{revtex}

\tightenlines
\newtheorem{theorem}{Theorem}
\newtheorem{acknowledgement}[theorem]{Acknowledgement}

\begin{document}
\title{Cosmic Strings Coupled with a Massless Scalar Field}
\author{O.Gurtug and I.Sakalli}
\address{Department of Physics, Eastern Mediterranean University G. Magosa, North\\
Cyprus, Mersin 10 - Turkey.\\
E-mail: ozay.gurtug@emu.edu.tr}
\date{\today }
\maketitle

\begin{abstract}
A scalar field generalization of Xanthopoulos's cylindrically symmetric
solutions of the vacuum Einstein equations is obtained. The obtained
solution preserves the properties of the Xanthopoulos solution, which are
regular on the axis, asymptotically flat and free from the curvature
singularities. The solution describes stable, infinite length of rotating
cosmic string interacting with gravitational and scalar waves.
\end{abstract}

\section{INTRODUCTION}

The first exact solution to the vacuum - Einstein equations in cylindrical
geometry is dated back to 1925 by Guido Beck [1]. However, the well known
solutions in this geometry belongs to Einstein and Rosen (ER) [2]. The
solution presented by ER represents first exact radiative solutions. The
solution is radiative because the waves carry away energy from the mass
located at the axis of symmetry. Furthermore, the solution obtained by ER
have two spacelike Killing vectors which are hypersurface orthogonal. The
most general vacuum solution describing cylindrical waves was studied
independently by Kompaneets and by Ehlers et al [3]. The cross polarized
version of ER waves was given by Halilsoy [4].

Chandrasekhar has constructed rather different type of formalism for
obtaining cylindrical waves with cross polarizations, similar to that used
for the discussion of the collision of plane gravitational waves [5]. The
main motivation for obtaining new cylindrical waves was to find applications
in general relativity and astrophysics. One of the application arena for
cylindrical spacetimes are cosmic strings. Cosmic strings are known to be
topological defects that formed during the cosmological phase transitions as
a result of spontaneous symmetry breaking of the Grand Unified Theory.
Cosmic strings would have important consequences for astrophysics. To say
the least, it is believed that the cosmic strings are responsible for galaxy
formation and gravitational lensing.

A long time ago, by employing the method of Chandrasekhar,\ Xanthopoulos has
obtained a family of three parameter and time dependent solutions both in
vacuum- Einstein [6] (hereafter paper I) and Einstein - Maxwell theory[7]
(hereafter paper II ). These solutions are cylindrically symmetric solutions
with remarkable properties that, they are regular on the axis,
asymptotically flat away from the axis and free from the curvature
singularites. One of the parameters of the solution measures the azimuthal
angle deficit that the space time exhibits and signals the presence of
infinite length of a straight cosmic string.

One of the solution given in paper II, is just the Einstein-Maxwell
extension of paper I. In that solution, it was shown that the deficit angle
near the axis or away from that axis is independent of the electromagnetic
parameter. For a particular value of deficit parameter $\alpha $, the
deficit angle vanishes and the solution represents propogation of
cylindirical gravitational and electromagnetic fields.

In this paper we present the Einstein-Scalar (ES) extension of paper I, with
the properties outlined in the third paragraph. The solution describes the
interaction of a spinning cosmic string with cylindrical gravitation and
scalar fields. Our solution is one-parameter generalization of paper I. In
addition to the parameters of paper I, we introduce a parameter $\beta $
that controls the intensity of the scalar field, such that for $\beta =0$
the solution reduces to the one given in paper I. In our analysis we have
shown that in contrast to the paper II, the deficit angle becomes dependent
on the scalar field parameter $\beta .$

The paper is organised as follows; in section II we review the solution
given in paper I. Section III deals with the construction of the
Einstein-Scalar solution. In section IV we discussed the physical properties
of the solution. We conclude the paper with a discussion in section V.

\section{REVIEW OF THE XANTHOPOULOS'S SOLUTION}

This solution represents a three parameters time-dependent cylindirically
symmetric solutions of the vacuum Einstein equations. The interesting
properties of the solution are,

(a) it is asymptotically flat.

(b) it admits a regular axis and

(c) it is free of curvature singularities and

(d) it exhibits an angle deficit in going around the axis and it is
interpreted as an infinitely long cosmic strings surrounded by gravitational
field.

Furthermore, the solution is Petrov type-D, and describes the propogation of
non-radiating gravitational waves. The non-radiating property implies the
stability of the cosmic string when it is interacting with gravitational
waves. The technique used by Xanthopoulos in obtaining the solution is the
one introduced by Chandrasekhar. In order to provide the regularity on the
axis and asymptotic flatness behavior, he adopts the prolate coordinate
system which is found very \ useful in the description of interacting plane
gravitational waves. The following metric is obtained in these prolate
coordinate system.

\bigskip 
\begin{equation}
ds^{2}=\alpha ^{2}X\left( \frac{d\tau ^{2}}{\Delta }-\frac{d\sigma ^{2}}{%
\delta }\right) -\frac{\Delta \delta X}{Y}d\varphi ^{2}-\frac{Y}{X}\left(
dz-q_{2}d\varphi \right) ^{2}
\end{equation}

where,

\begin{eqnarray}
X &=&\left( 1-p\tau \right) ^{2}+q^{2}\sigma ^{2} \\
Y &=&p^{2}\tau ^{2}+q^{2}\sigma ^{2}-1=p^{2}\Delta +q^{2}\delta  \nonumber \\
q_{2} &=&\frac{2q\delta \left( 1-p\tau \right) }{pY}  \nonumber \\
\Delta &=&\tau ^{2}+1  \nonumber \\
\delta &=&\sigma ^{2}-1  \nonumber
\end{eqnarray}

such that $\tau \in \Re $ and $\sigma \geq 1.$ The parameters $\alpha $,$p$
and $q$ are constants with $q^{2}-p^{2}=1$. Using the following
transformation

\[
\omega =\sqrt{\Delta \delta }\text{ , \ \ \ \ \ \ \ }t=\tau \sigma 
\]

line element $\left( 1\right) $ transforms into cylindirical coordinates

\begin{equation}
ds^{2}=\frac{\alpha ^{2}X}{\tau ^{2}+\sigma ^{2}}\left( dt^{2}-d\omega
^{2}\right) -\frac{\omega ^{2}X}{Y}d\varphi ^{2}-\frac{Y}{X}\left(
dz-q_{2}d\varphi \right) ^{2}
\end{equation}

with

\[
2\tau ^{2}=\sqrt{D}+t^{2}-\omega ^{2}-1,\text{ \ \ \ \ \ \ \ \ \ \ }2\sigma
^{2}=\sqrt{D}-t^{2}+\omega ^{2}+1 
\]

where

\[
\sqrt{D}=\left( \omega ^{2}-t^{2}+1\right) ^{2}+4t^{2}\geqslant 0. 
\]

Therefore the solution is analysed in the cylindirical coordinates while the
field equations are solved in some other coordinates. With reference to the
detailed analysis in paper I, the behavior of the metric functions near the
axis $\left( \omega \rightarrow 0^{+}\right) $ and asymptotically $\left(
\omega \rightarrow +\infty \right) $ are obtained by using the following
relations.

Near the axis $\omega \ll \left| t\right| $ , $t=$finite

\begin{eqnarray}
\tau &\simeq &t-\frac{\omega ^{2}t}{2\left( 1+t^{2}\right) }+O\left( \omega
^{4}\right) \\
\sigma &\simeq &1+\frac{\omega ^{2}}{2\left( 1+t^{2}\right) }+O\left( \omega
^{4}\right)  \nonumber
\end{eqnarray}

Asymptotically $\omega \gg \left| t\right| $

\begin{eqnarray}
\tau &\simeq &\frac{t}{\omega }+\frac{t\left( t^{2}-1\right) }{2\omega ^{3}}%
+O\left( \omega ^{-4}\right) \\
\sigma &\simeq &\omega +\frac{\left( 1-t^{2}\right) }{2\omega }+O\left(
\omega ^{-2}\right)  \nonumber
\end{eqnarray}

In paper I, it was shown that the deficit angle which indicates the presence
of cosmic strings obtained as 
\begin{equation}
\delta \varphi =2\pi \left( 1-e^{-C}\right)
\end{equation}

where, $C$ is the $C$-energy and it was also shown \ that this energy is
non-radiating. The absence of radiation can also be justified by calculating
the Weyl scalars, which are all vanishing in the asymptotic region.

\section{CONSTRUCTION OF THE EINSTEIN-SCALAR SOLUTION}

Cylindirical gravitational waves with cross polarization are described in
general by the line element of Jordan- Ehlers-Kundt- Kompaneetz [3] as

\begin{equation}
ds^{2}=e^{2\left( \gamma -\psi \right) }\left( dt^{2}-d\omega ^{2}\right)
-\omega \left[ \omega e^{-2\psi }d\varphi ^{2}+\frac{e^{2\psi }}{\omega }%
\left( dz+q_{2}d\varphi \right) ^{2}\right]
\end{equation}

In order to couple massless scalar field to the system we write the
Lagrangian density as follows

\begin{equation}
L=\left( \lambda _{\omega }\gamma _{\omega }-\lambda _{t}\gamma _{t}\right)
-\lambda \left[ \psi _{\omega }^{2}-\psi _{t}^{2}+2\left( \phi _{\omega
}^{2}-\phi _{t}^{2}\right) \right] -\frac{e^{4\psi }}{4\lambda }\left(
q_{2\omega }^{2}-q_{2t}^{2}\right)
\end{equation}

where \ $\phi $ is the scalar field and $\gamma =\gamma \left( \omega
,t\right) ,\psi =\psi \left( \omega ,t\right) ,\phi =\phi \left( \omega
,t\right) ,q_{2}=q_{2}\left( \omega ,t\right) $ and $\lambda $ is a
coordinate condition and for the present problem we choose it as $\lambda
=\omega $.

The Einstein- Scalar field equations, which are obtained by varying the
Lagrangian described \ in equation $\left( 8\right) $, are

\begin{equation}
\psi _{tt}-\frac{\psi _{\omega }}{\omega }-\psi _{\omega \omega }=\frac{%
e^{4\psi }}{2\omega ^{2}}\left( q_{2t}^{2}-q_{2\omega }^{2}\right)
\end{equation}

\begin{equation}
q_{2tt}+\frac{q_{2\omega }}{\omega }-q_{2\omega \omega }=4\left( q_{2\omega
}\psi _{\omega }-q_{2t}\psi _{t}\right)
\end{equation}

\begin{equation}
\gamma _{\omega }=\omega \left( \psi _{t}^{2}+\psi _{\omega }^{2}\right) +%
\frac{e^{4\psi }}{4\omega }\left( q_{2t}^{2}+q_{2\omega }^{2}\right)
+2\omega \left( \phi _{\omega }^{2}+\phi _{t}^{2}\right)
\end{equation}

\begin{equation}
\gamma _{t}=2\omega \psi _{\omega }\psi _{t}+\frac{e^{4\psi }}{2\omega }%
q_{2t}q_{2\omega }+4\omega \phi _{\omega }\phi _{t}
\end{equation}

\begin{equation}
\phi _{\omega \omega }+\frac{\phi _{\omega }}{\omega }-\phi _{tt}=0
\end{equation}

The solution to these equations are obtained by employing the formalism of
Chandrasekhar. Hence, we prefer to use the notation of Chandrasekhar [5].

If we set

\begin{equation}
\nu =\gamma -\psi \text{ \ \ \ and \ \ \ \ \ \ \ }\chi =\omega e^{-2\psi }
\end{equation}

Line element $\left( 7\right) $ becomes

\begin{equation}
ds^{2}=e^{2\nu }\left( dt^{2}-d\omega ^{2}\right) -\omega \left[ \chi
d\varphi ^{2}+\chi ^{-1}\left( dz-q_{2}d\varphi \right) ^{2}\right]
\end{equation}

where $\frac{\partial }{\partial z}$ and \ $\frac{\partial }{\partial
\varphi }$ are the axial and azimuthal Killing fields and $\nu ,\chi $ and $%
q_{2}$ are functions of $\omega $ and $t.$ Using equation $\left( 14\right) $
in the field equations $\left( 11\right) $ and $\left( 12\right) $ they
transform into

\begin{equation}
4\nu _{\omega }=-\frac{1}{\omega }+\frac{\omega }{\chi }\left[ \chi
_{t}^{2}+\chi _{\omega }^{2}+q_{2t}^{2}+q_{2\omega }^{2}\right] +8\omega
\left( \phi _{\omega }^{2}+\phi _{t}^{2}\right)
\end{equation}

\begin{equation}
\nu _{t}=\frac{\omega }{2\chi ^{2}}\left[ \chi _{t}\chi _{\omega
}+q_{2t}q_{2\omega }\right] +4\omega \phi _{\omega }\phi _{t}
\end{equation}

respectively. In terms of the Ernst potentials ($\Psi $ and $\Phi $) defined
in paper I equations $\left( 16\right) $ and $\left( 17\right) $ become

\begin{equation}
\left( \nu +\ln \sqrt{\Psi }\right) _{t}=\frac{\omega }{2\Psi ^{2}}\left(
\Psi _{t}\Psi _{\omega }+\Phi _{t}\Phi _{\omega }\right) +4\omega \phi
_{\omega }\phi _{t}
\end{equation}

\begin{equation}
\left( \nu +\ln \sqrt{\Psi }\right) _{\omega }=\frac{\omega }{4\Psi ^{2}}%
\left( \Psi _{t}^{2}+\Psi _{\omega }^{2}+\Phi _{t}^{2}+\Phi _{\omega
}^{2}\right) +2\omega \left( \phi _{\omega }^{2}+\phi _{t}^{2}\right)
\end{equation}

The scalar field $\phi $ is coupled to the system by shifting the metric
function $\nu $ in accordance with [8]

\begin{equation}
\nu =\nu _{0}+\Gamma
\end{equation}

where

\begin{eqnarray}
\Gamma _{t} &=&4\omega \phi _{\omega }\phi _{t} \\
\Gamma _{\omega } &=&2\omega \left( \phi _{\omega }^{2}+\phi _{t}^{2}\right)
\nonumber
\end{eqnarray}

and $\nu _{0},\chi _{0}$ and $q_{20}$ satisfy the vacuum Einstein equations
and $\Gamma $ is the additional metric function that arises due to the
presence of the scalar field. Integrability condition for the equations $%
\left( 21\right) $ imposes the massless scalar field equation $\left(
13\right) $ as a constraint condition from which we can generate a large
class of ES solution. The solution to the ES field equations will be
obtained in the prolate coordinates as a requirement of the formalism of
Chandrasekhar. If we set,

\[
\omega =\sqrt{\Delta \delta }\text{, \ \ \ \ }t=\tau \sigma \text{, \ \ \ \
\ \ }\Delta =\tau ^{2}+1\text{, \ \ \ \ }\delta =\sigma ^{2}-1 
\]

we can express the theory of cylindirical gravitational and scalar waves in
the $\left( \tau ,\sigma \right) $ coordinates whose range are

\[
\tau \in \Re \text{ \ \ \ \ \ \ \ \ \ \ \ \ \ \ \ \ \ \ }\sigma \geqslant 1 
\]

note that $\sigma =1$ corresponds to the axis $\omega =0$. We find \ that
line element $\left( 15\right) $ becomes

\begin{equation}
ds^{2}=\left( \tau ^{2}+\sigma ^{2}\right) e^{2\left( \nu _{0}+\Gamma
\right) }\left[ \frac{d\tau ^{2}}{\Delta }-\frac{d\sigma ^{2}}{\delta }%
\right] -\sqrt{\Delta \delta }\left[ \chi _{0}d\varphi ^{2}+\chi
_{0}^{-1}\left( dz-q_{20}d\varphi \right) ^{2}\right]
\end{equation}

In this paper we wish to couple a massless scalar field to the solution
presented in paper I. Hence as an Einstein-vacuum solution $\nu _{0},\chi
_{0}$ and $q_{20}$ , we shall make use of the solution obtained in paper I
that describes the cylindirical gravitational waves with interesting
properties discussed. Therefore the resulting metric that describes ES
solution is given by

\begin{equation}
ds^{2}=\frac{\alpha ^{2}X}{\tau ^{2}+\sigma ^{2}}e^{2\Gamma }\left(
dt^{2}-d\omega ^{2}\right) -\frac{\omega ^{2}X}{Y}d\varphi ^{2}-\frac{Y}{X}%
\left( dz-q_{2}d\varphi \right) ^{2}
\end{equation}

where $X,Y$ and $q_{2}$ are given in equation $\left( 2\right) .$ In terms
of the new coordinates equation $\left( 21\right) $ becomes

\begin{eqnarray}
\Gamma _{\sigma } &=&\frac{2\Delta }{\tau ^{2}+\sigma ^{2}}\left[ \sigma
\delta \phi _{\sigma }^{2}+\sigma \Delta \phi _{\tau }^{2}-2\tau \delta \phi
_{\sigma }\phi _{\tau }\right] \\
\Gamma _{\tau } &=&\frac{2\delta }{\tau ^{2}+\sigma ^{2}}\left[ 2\sigma
\Delta \phi _{\sigma }\phi _{\tau }-\tau \delta \phi _{\sigma }^{2}-\tau
\Delta \phi _{\tau }^{2}\right]  \nonumber
\end{eqnarray}

The massless scalar field equation in terms of \ $\left( \tau ,\sigma
\right) $ may be written as

\begin{equation}
\left( \Delta \phi _{\tau }\right) _{\tau }-\left( \delta \phi _{\sigma
}\right) _{\sigma }=0
\end{equation}

This equation has many solutions. Among the others we wish to adopt as a
scalar field, Bonnor's non-singular cylindirical wave solutions found long
time ago [9]. The interesting property of this solution is that, it is
regular near the axis and vanishes asymptotically. In terms of $\left( \tau
,\sigma \right) $ coordinates the scalar field is given by

\begin{equation}
\phi \left( \tau ,\sigma \right) =\frac{\beta \sigma }{\tau ^{2}+\sigma ^{2}}
\end{equation}

where $\beta $ is a constant parameter measuring the intensity of the scalar
wave. Using equation $\left( 24\right) $ the additional metric function $%
\Gamma $ is found as

\[
\Gamma \left( \tau ,\sigma \right) =\frac{\beta ^{2}\Delta }{\tau
^{2}+\sigma ^{2}}\left[ \frac{4\tau ^{4}\Delta }{\left( \tau ^{2}+\sigma
^{2}\right) ^{3}}-\frac{4\tau ^{2}\left( 1+2\tau ^{2}\right) }{\left( \tau
^{2}+\sigma ^{2}\right) ^{2}}+\frac{1+9\tau ^{2}}{2\left( \tau ^{2}+\sigma
^{2}\right) }-1\right] 
\]

Extending this to the Einstein-Maxwell-Scalar version of paper II is
straightforward.

\section{PHYSICAL PROPERTIES OF THE SOLUTION}

\subsection{Boundary Conditions}

In order to interpret the present solution physically acceptable, we should
impose certain conditions on the behavior of the metric functions near the
symmetry axis and asymptotically far away from the axis. These conditions,
given in paper I, are as follows:

(a) The symmetry axis should be regular when $\omega \rightarrow 0$. This
means that, the squared norm of the rotational (or azimuthal) Killing field $%
\left( \frac{\partial }{\partial \varphi }\right) $

\begin{equation}
\left| \frac{\partial }{\partial \varphi }\right| ^{2}=-\omega \frac{\left(
\chi ^{2}+q_{2}^{2}\right) }{\chi }
\end{equation}

should approach zero like $\omega ^{2}$ near $\omega =0$. This condition
quarantees that the axis is regular.

(b) Asymptotically far away from the symmetry axis, i.e. in all directions
such that $\omega \rightarrow \infty $, the solution should become
asymptotically flat. It should be noted that this solution is not
asypmtotically simple in the sense of Penrose. The reason is the spatial
infinity in the direction $z\rightarrow \infty $ cannot be carried out,
because, $\frac{\partial }{\partial z}$ is a Killing vector and the metric
functions are independent of $z$.

\subsection{Behaviour Near The Axis : $\protect\omega \ll t$ $\left( \protect%
\omega \rightarrow 0\right) $}

Using the expressions given in equation $\left( 4\right) $ we find the
behavior of the metric functions near the axis$\left( \omega \rightarrow
0^{+}\right) $ as follows

\begin{eqnarray}
1-p\tau &\simeq &1-pt+O\left( \omega ^{2}\right) \\
\delta &\simeq &\frac{\omega ^{2}}{t^{2}+1}+O\left( \omega ^{4}\right) 
\nonumber \\
X &\simeq &\left[ \left( 1-pt\right) ^{2}+q^{2}\right] +O\left( \omega
^{2}\right)  \nonumber \\
Y &\simeq &p^{2}\left( 1+t^{2}\right) +O\left( \omega ^{2}\right)  \nonumber
\\
\tau ^{2}+\sigma ^{2} &\simeq &1+t^{2}+O\left( \omega ^{2}\right)  \nonumber
\\
e^{2\Gamma } &=&e^{-\beta ^{2}}  \nonumber
\end{eqnarray}

so that:

\begin{eqnarray}
\chi &\simeq &\frac{A}{p^{2}}\omega +O\left( \omega ^{3}\right) \\
q_{2} &\simeq &\frac{2q\left( 1-pt\right) }{p^{3}\left( 1+t^{2}\right) ^{2}}%
\omega ^{2}+O\left( \omega ^{3}\right)  \nonumber \\
\frac{\chi ^{2}+q_{2}^{2}}{\chi } &\simeq &\frac{A}{p^{2}}\omega +O\left(
\omega ^{2}\right)  \nonumber
\end{eqnarray}

where

\[
A=\frac{\left( 1-pt\right) ^{2}+q^{2}}{1+t^{2}} 
\]

The norms of the two Killing fields behave like

\bigskip 
\begin{eqnarray}
\left| \frac{\partial }{\partial \varphi }\right| ^{2} &=&-\omega \frac{\chi
^{2}+q_{2}^{2}}{\chi }\simeq -\frac{A}{p^{2}}\omega ^{2}+O\left( \omega
^{3}\right) \\
\left| \frac{\partial }{\partial z}\right| ^{2} &=&-\frac{\omega }{\chi }%
\simeq \frac{p^{2}}{A}+O\left( \omega \right)  \nonumber
\end{eqnarray}

We show that in the limit $\omega \rightarrow 0^{+}$ is a regular surface of
the spacetime. Note that, the coupling of the scalar field does not play a
role in the behavior of the two Killing fields. Using expressions (28 -29 ),
we find the metric near the axis;

\begin{equation}
ds^{2}\simeq \alpha ^{2}Ae^{-\beta ^{2}}\left[ dt^{2}-d\omega ^{2}-\frac{%
\omega ^{2}e^{\beta ^{2}}}{p^{2}\alpha ^{2}}d\varphi ^{2}\right] -\frac{p^{2}%
}{A}dz^{2}+\frac{4q\left( 1-pt\right) \omega ^{2}}{p\left( 1+t^{2}\right)
^{2}A}dzd\varphi +O\left( \omega ^{3}\right)
\end{equation}

\subsection{Asymptotic Behaviour : $\protect\omega \gg \left| t\right| $ \ $%
\left( \protect\omega \rightarrow \infty \right) $}

Using the expressions (5)$,$ we find that

\begin{eqnarray}
X &\simeq &q^{2}\omega ^{2}+\left( 1+q^{2}-q^{2}t^{2}\right) +O\left( \omega
^{-1}\right) \\
Y &\simeq &q^{2}\omega ^{2}+\left( p^{2}-q^{2}t^{2}\right) +O\left( \omega
^{-1}\right)  \nonumber \\
\delta &\simeq &\omega ^{2}-t^{2}+O(\omega ^{-1})  \nonumber \\
q_{2} &\simeq &\frac{2}{pq}\left[ 1-\frac{pt}{\omega }\right] +O\left(
\omega ^{-2}\right)  \nonumber \\
\tau ^{2}+\sigma ^{2} &\simeq &\omega ^{2}+\left( 1-t^{2}\right) +O\left(
\omega ^{-1}\right)  \nonumber \\
\Gamma &\simeq &0  \nonumber
\end{eqnarray}

Line element (23) , asymptotically becomes;

\begin{equation}
ds^{2}\simeq \alpha ^{2}q^{2}\left[ dt^{2}-d\omega ^{2}-\frac{\omega ^{2}}{%
\alpha ^{2}q^{2}}d\varphi ^{2}\right] -\left( dz-\frac{2}{pq}d\varphi
\right) ^{2}
\end{equation}

by letting \ $z\rightarrow \widetilde{z}=z-\frac{2}{pq}\varphi ,$ the last
term in the metric is changed and metric \ (23 ) becomes asymptotically flat.

\subsection{Existence of Cosmic Strings}

Comparing the two metrics obtained for near the axis $\left( \omega
\rightarrow 0\right) $ and asymptotic case $\left( \omega \rightarrow \infty
\right) $, we observe that exact metric (23) allows an angle deficit
measured by $\frac{e^{\beta ^{2}}}{\alpha ^{2}p^{2}}$ near the axis and $%
\frac{1}{\alpha ^{2}q^{2}}$ asymptotically far away from the axis. This can
be shown as follows.

For any metric in the form of

\begin{equation}
ds^{2}=f\left[ d\omega ^{2}+k^{2}\omega ^{2}d\varphi ^{2}\right] +.....
\end{equation}

with constant $k$ and $f$ independent of $\omega $, the angle deficit near
the axis $\left( \omega \rightarrow 0\right) $ or asymptotically $\left(
\omega \rightarrow \infty \right) ,$ can be obtained from the definition:

\begin{equation}
\delta \varphi =2\pi -%
\mathrel{\mathop{\lim }\limits_{\omega \rightarrow 0(\infty )}}%
\frac{\int_{0}^{2\pi }\sqrt{g_{\varphi \varphi }}d\varphi }{\int_{0}^{\omega
}\sqrt{g_{\omega \omega }}d\omega }
\end{equation}

Using the above definition, we find the deficit angle as,

\begin{equation}
\delta \varphi =2\pi \left( 1-k\right)
\end{equation}

Exact metric (23) displays a conical singularity on the axis, measured by
the angle deficit, which signals the existence of \ the cosmic string on the
axis.

\begin{equation}
\left( \delta \varphi \right) _{ax}=2\pi \left( 1-\frac{e^{\frac{\beta ^{2}}{%
2}}}{\left| \alpha \right| \left| p\right| }\right)
\end{equation}

For $\alpha =e^{\frac{\beta ^{2}}{2}}\left| p\right| ^{-1}$ it removes the
angle deficit and implies the absence of the cosmic string. The axis becomes
locally flat, and the metric (23) represents the propogation of coupled
cylindirical gravitational and scalar waves.

In references $\left[ 10,11,12,13\right] $ it has been shown that the mass \
per unit length of the string \ $\mu _{0}$ is related to the angle deficit by

\begin{equation}
8\pi \mu _{0}=\delta \varphi
\end{equation}

Hence, the mass density $\mu _{0}$ for the present paper becomes

\begin{equation}
\mu _{0}=\frac{1}{4}\left( 1-\frac{e^{\frac{\beta ^{2}}{2}}}{\left| \alpha
\right| \left| p\right| }\right)
\end{equation}

Note that the mass per unit length $\mu _{0}$ is constant, does not depend
on time $t.$

Asymptotically the spacetime exhibits an angle deficit given by

\begin{equation}
(\delta \varphi )_{asy}=2\pi \left( 1-\frac{1}{\left| \alpha \right| \left|
q\right| }\right)
\end{equation}

For the value of $\alpha =e^{\frac{\beta ^{2}}{2}}\left| p\right| ^{-1}$
eliminates the angle deficit near the axis while asymptotically the angle
deficit becomes

\begin{equation}
\left( \delta \varphi \right) _{asy}=2\pi \left( 1-\frac{p}{qe^{\frac{\beta
^{2}}{2}}}\right)
\end{equation}

Since \ $q^{2}-p^{2}=1$ , this implies that $\left| q\right| >\left|
p\right| $ and gives a larger angle deficit compared to the result obtained
\ in paper I.

The choice $\alpha =\left| q\right| ^{-1}$, removes the asymptotic angle
deficit and the angle deficit near the axis becomes

\begin{equation}
\left( \delta \phi \right) _{ax}=2\pi \left( 1-\frac{qe^{\frac{\beta ^{2}}{2}%
}}{p}\right)
\end{equation}

This choice causes a larger negative angle surplus near the axis compared
with the paper I. The choice $\alpha >e^{\frac{\beta ^{2}}{2}}\left|
p\right| ^{-1}$ imposes deficit angle both in asymptotically and near the
axis. Note that; for this particular case asymptotic angle deficit is
greater than the angle deficit near the axis. This property is also
encountered in the analysis of paper I and paper II. It has been shown in
paper II that, the angle deficit near the axis and asymptotically, are
independent of electromagnetic parameter. However, in our case the angle
deficit near the axis depends on the constant parameter $\beta $, which
measures the intensity of the scalar wave but asymptotically the angle
deficit is independent of the scalar field. This is natural and expected
because the intensity of the scalar field is maximum near the axis and
approaches zero for far away from the axis. Hence, in contrast to the
electromagnetic case (paper II), intervening gravitational and scalar waves
contribute to the angle deficit.

\subsection{Discussion of \ Energy}

One of the important elements in cylindirically symmetric systems is the $C$%
-energy introduced by Thorne [14]. This $C$-energy represents the total
gravitational and scalar energy per unit length between $\omega =0$ and $%
\omega $ at any time $t$. The $C$-energy in the present paper is described
by the quantity.

\begin{equation}
C=\nu +\ln \sqrt{\Psi }+\Gamma
\end{equation}

which is equivalent to

\begin{eqnarray}
C &=&\frac{1}{2}\ln \left[ \frac{\alpha ^{2}\left( p^{2}\tau
^{2}+q^{2}\sigma ^{2}-1\right) }{\tau ^{2}+\sigma ^{2}}\right] + \\
&&\frac{\beta ^{2}\Delta }{\tau ^{2}+\sigma ^{2}}\left[ \frac{4\tau
^{4}\Delta }{\left( \tau ^{2}+\sigma ^{2}\right) ^{3}}-\frac{4\tau
^{2}\left( 1+2\tau ^{4}\right) }{\left( \tau ^{2}+\sigma ^{2}\right) ^{2}}+%
\frac{1+9\tau ^{2}}{2\left( \tau ^{2}+\sigma ^{2}\right) }-1\right] 
\nonumber
\end{eqnarray}

Near the axis (when $\sigma =1$ or $\omega \rightarrow 0$), equation ( 44)
gives

\begin{equation}
C\simeq \ln \left| \alpha p\right| -\frac{\beta ^{2}}{2}\text{ \ \ \ \ \ \ \
\ \ \ \ \ \ \ }\left( \omega \rightarrow 0\right)
\end{equation}

while asymptotically it becomes

\begin{equation}
C\simeq \ln \left| \alpha q\right| +\ln \left| 1-\frac{1}{2q^{2}\omega ^{2}}%
\right| \text{ \ \ \ \ \ \ \ \ \ \ \ \ \ \ \ \ \ \ }\left( w\rightarrow
\infty \right)
\end{equation}

Using equation (43) in equations $\left( 18\right) $ and $\left( 19\right) $
we obtained

\begin{equation}
C_{t}=\frac{\omega }{2\Psi ^{2}}\left( \Psi _{t}\Psi _{\omega }+\Phi
_{t}\Phi _{\omega }\right) +4\omega \phi _{\omega }\phi _{t}
\end{equation}

\begin{equation}
C_{\omega }=\frac{\omega }{4\Psi ^{2}}\left( \Psi _{t}^{2}+\Phi
_{t}^{2}+\Psi _{\omega }^{2}+\Phi _{\omega }^{2}\right) +2\omega \left( \phi
_{\omega }^{2}+\phi _{t}^{2}\right)
\end{equation}

We introduce the null coordinates

\begin{eqnarray}
u &=&t+\omega \\
v &=&t-\omega  \nonumber
\end{eqnarray}

so that future null infinity corresponds to $u\rightarrow \infty $ with
finite $v$ while past null infinity corresponds to $v\rightarrow -\infty $
with finite $u$. Equations (18) and (19) take the form \ \ \ \ 

\begin{equation}
C_{u}=\frac{1}{2}\left( C_{t}+C_{\omega }\right) >0
\end{equation}

\begin{equation}
C_{v}=\frac{1}{2}\left( C_{t}-C_{\omega }\right)
\end{equation}

where equation (51) measures the rate of radiation of the $C$ energy . For
the present paper, equation (51) \ interms of $(\tau ,\sigma )$ coordinates
takes the form

\begin{equation}
C_{v}=-\frac{\left( \sigma \sqrt{\Delta }+\tau \sqrt{\delta }\right) ^{2}}{%
\left( \tau ^{2}+\sigma ^{2}\right) ^{2}}\left\{ \frac{\sqrt{\Delta \delta }%
}{2Y}+\frac{\omega \beta ^{2}}{\left( \tau ^{2}+\sigma ^{2}\right) ^{4}}%
\left[ \sqrt{\delta }\left( \tau ^{2}-\sigma ^{2}\right) +2\tau \sigma \sqrt{%
\Delta }\right] ^{2}\right\}
\end{equation}

Using the following asymptotic behavior

\begin{eqnarray}
\sqrt{\Delta } &\rightarrow &1\text{ \ \ \ \ \ \ \ \ \ \ \ \ \ \ \ \ \ }%
\sqrt{\delta }\rightarrow \omega \text{ \ \ \ \ \ \ \ \ \ \ \ \ \ \ \ \ \ \
\ \ \ \ \ \ }\tau ^{2}+\sigma ^{2}\simeq \omega ^{2}-t^{2}+1 \\
\tau ^{2}-\sigma ^{2} &\simeq &-\omega ^{2}+t^{2}-1\text{ \ \ \ \ \ \ \ \ \
\ \ \ \ \ \ \ \ }Y\simeq q^{2}\omega ^{2}+\left( p^{2}-q^{2}t^{2}\right) 
\nonumber \\
\sigma &\simeq &\omega +\frac{1-t^{2}}{2\omega }\text{ \ \ \ \ \ \ \ \ \ \ \
\ \ \ \ \ \ \ \ \ \ \ \ \ \ \ \ }\tau \simeq \frac{t}{\omega }+\frac{t\left(
t^{2}-1\right) }{2\omega ^{3}}  \nonumber
\end{eqnarray}

we obtain that

\begin{equation}
C_{v}\simeq -\frac{1}{2\omega ^{3}}\left( 1+\frac{2t}{\omega }\right) \left( 
\frac{1}{q^{2}}+2\beta ^{2}\right) +O\left( \omega ^{-5}\right)
\end{equation}

It is clear to observe that $%
\mathrel{\mathop{\lim }\limits_{\omega \rightarrow \infty }}%
C_{v}=0$ and the solution is non-radiating

\subsection{The Weyl and Ricci \ Scalars}

The description of the solution in terms of the Weyl and Ricci scalars are
obtained by Newman-Penrose formalism. We introduce new coordinates $\left(
\theta ,\psi \right) $ by

\begin{eqnarray}
\tau &=&\sinh \psi \\
\sigma &=&\cosh \theta  \nonumber
\end{eqnarray}

The line element $\left( 23\right) $ describing ES solution takes the form

\bigskip 
\begin{equation}
ds^{2}=U^{2}\left( d\psi ^{2}-d\theta ^{2}\right) -\frac{V^{2}}{%
1-\varepsilon \varepsilon ^{\ast }}\left[ \left( 1-\varepsilon \right)
dz+i\left( 1+\varepsilon \right) d\varphi \right]
\end{equation}

where

\[
U^{2}=\alpha ^{2}Xe^{2\Gamma } 
\]

and a suitable null basis for the metric is

\begin{eqnarray}
l &=&\frac{U}{\sqrt{2}}\left( d\psi -d\theta \right) \\
n &=&\frac{U}{\sqrt{2}}\left( d\psi +d\theta \right)  \nonumber \\
m &=&-\frac{1}{\sqrt{2}}\left( VL_{-}^{\ast }dz-iVL_{+}^{\ast }d\varphi
\right)  \nonumber
\end{eqnarray}

Here

\[
L_{\pm }=\frac{1\pm \varepsilon }{\sqrt{1-\left| \varepsilon \right| ^{2}}} 
\]

\[
V^{2}=\cosh \psi \sinh \theta 
\]

\[
\varepsilon =\frac{Z-1}{Z+1}\text{ \ \ \ \ \ \ where\ \ \ \ \ \ \ \ \ \ \ \
\ \ \ }Z=\chi +iq_{2} 
\]

and $\ast $ denotes the complex conjugate. The exact form of the Weyl and
Ricci scalars are too long and complicated. Hence, we prefer to give their
asymptotic forms

\begin{equation}
\psi _{2}\simeq -\frac{i}{2\alpha ^{2}q^{3}\omega ^{3}}
\end{equation}

\begin{equation}
\psi _{0}\simeq -\frac{3i}{2\alpha ^{2}q^{3}\omega ^{3}}+\frac{\beta ^{2}}{%
2\alpha ^{2}q^{2}\omega ^{4}}+O\left( \omega ^{-5}\right)
\end{equation}

\begin{equation}
\psi _{4}\simeq -\frac{3i}{2\alpha ^{2}q^{3}\omega ^{3}}+\frac{\beta ^{2}}{%
2\alpha ^{2}q^{2}\omega ^{4}}+O\left( \omega ^{-5}\right)
\end{equation}

\begin{equation}
\phi _{00}\simeq \frac{\beta ^{2}}{2\alpha ^{2}q^{2}\omega ^{4}}\left( 1-%
\frac{1}{\omega ^{2}}\right) +O\left( \omega ^{-7}\right)
\end{equation}

\begin{equation}
\phi _{22}\simeq \frac{\beta ^{2}}{2\alpha ^{2}q^{2}\omega ^{4}}\left( 1-%
\frac{1}{\omega ^{2}}\right) +O\left( \omega ^{-7}\right)
\end{equation}

\begin{equation}
\phi _{11}\simeq \frac{\beta ^{2}}{2\alpha ^{2}q^{2}\omega ^{4}}\left( 1+%
\frac{q-11}{2\omega ^{2}}\right) +O\left( \omega ^{-7}\right)
\end{equation}

\begin{equation}
\Lambda \simeq -\frac{\beta ^{2}}{6\alpha ^{2}q^{2}\omega ^{4}}\left( 1+%
\frac{q-11}{2\omega ^{2}}\right) +O\left( \omega ^{-7}\right)
\end{equation}

The asymptotic dying-off of the Weyl (like $\omega ^{-3}$ ) and Ricci ( like 
$\omega ^{-4}$) scalars indicates that the present solution is
non-radiating. This is totally in agreement with the $C$-energy discussion.

The non-radiating character of the present solution indicates the stability
of the cosmic string in the presence of a scalar field as well.

It should be noted that the inclusion of scalar field changes the geometric
intrepretation of the resulting spacetime from Petrov type-D to Petrov
type-I.

\section{DISCUSSION}

In this paper, we have given the scalar field generalization of the vacuum
Einstein solution of paper I, that describes the interaction of rotating
cosmic strings with gravitational waves. The obtained solution is a kind of
special solution that preserves the properties of the background solution;
that is, regular on the axis, asymptotically flat and free from the
curvature singularities. It is shown that the presence of a scalar field
contributes to the angle deficit near the axis. However in the asymptotic
case, the angle deficit is produced completely from the gravitational waves.
Our analysis shows that, the effect of the scalar field tends to increase
the angle deficit. We also note that the inclusion of the scalar field did
not change the stability character of the cosmic string. This is not
surprising because the character of the inserted scalar field is
well-behaved and asymptotically vanished. Addition of a different scalar
field which does not behave well will change all these nice features
completely.

\begin{acknowledgement}
We would like to thank Mustafa Halilsoy for discussions.
\end{acknowledgement}

\end{document}